\documentclass[11pt,a4paper]{desyproc}

\begin{document}
\title{\vspace*{-1cm}\normalsize \hfill\hspace*{11cm}\mbox{IPPP/10/89; DCPT/10/178}\hfill\mbox{}\\[1.5cm]
Spectroscopic Bounds on New Physics}

\author{{\slshape Joerg Jaeckel, Sabyasachi Roy}\\[1ex]
Institute for Particle Physics Phenomenology, Durham University, Durham DH1 3LE, United Kingdom\\}



\maketitle

\begin{abstract}
We use atomic spectra to extend pure Coulomb's law tests to larger masses. We interpret these results in terms of constraints for hidden sector photons. With existing data the bounds for hidden photons are not improved. However we find that our atomic spectra bounds are an especially clean and model-independent complement to existing ones from other methods. We also show that data from future tests of true muonium and muonic atoms could produce atomic spectra bounds which probe untested parameter space.
\end{abstract}

\section{Introduction}
\label{Introduction}

We use atomic spectroscopy of ordinary and exotic atoms to test Coulomb's law with high precision on atomic length scales~\cite{Popovontheexp, Karshenboim:2010cg, Karshenboim:2010ck}.
This in turn allows us to constrain new particles such as hidden photons~\cite{Popovontheexp, Okun, Pospelov:2008zw} which arise naturally in a variety of extensions of the standard model~\cite{Okun,Holdom,Dienes1997104, Abel:2003ue,  Abel:2006qt, Abel:2008ai,Goodsell:2009xc} (see also \cite{Jaeckel:2010ni} for a review)\footnote{Note that we can also produce bounds for minicharged particles \cite{Jaeckel:2010xx,Gluck:2007ia,Jaeckel:2009dh}. However they turn out to be relatively weak.}\footnote{Spectroscopy can also constrain Unparticles (see e.g.~\cite{Thalapillil:2009ch}).}. 

Hidden photons cause a deviation from Coulomb's law,
\begin{equation}
V(r)=-\frac{Z \alpha}{r} (1 + e^{-m_{\gamma^{\prime}} r} \chi^2) 
\label{Vintro}
\end{equation}
where $m_{\gamma^{\prime}}$ is the mass of the hidden photon and $\chi$ is the kinetic mixing~\cite{Holdom}. Note that independent of the particle interpretation, our bounds can more generally constrain deviations from Coulomb's law by a Yukawa type potential.

In the small and large mass limits we recover the original $\frac{1}{r}$ form potential. It is only in the intermediate mass regions that we expect to see measurable deviations to Coulomb's Law. Hence we expect our bounds to drop off at low and high energies.

\section{Spectroscopic bounds}
\label{spectroscopicbounds}
We adapt the method presented in Ref.~\cite{Gluck:2007ia}, where the Lamb shift in atomic hydrogen is used to bound minicharged particles. 

At first order in perturbation theory the energy shift of a state $|\psi_{n}\rangle$ is given by
\begin{equation}
\label{1st order}
\delta E_{n}^{(1)} = \langle \psi_{n} \mid H' \mid \psi_{n} \rangle= \langle \psi_{n} \mid \delta V \mid \psi_{n} \rangle .
\end{equation}

We then impose that $\delta E^{(1)}_{n}$ must be smaller than the uncertainty in the transition\footnote{We conservatively estimate the ``uncertainty'' by adding absolute values of the experimental and theoretical errors (see \cite{Jaeckel:2010xx}).}\footnote{Charge radii of nuclei are a major source of uncertainty. These radii must be determined from an independent source. Moreover, to avoid even partial degeneracies (which weaken the bound at short length scales), the determination of the radius should be obtained at high momentum transfer. Hence we use electron scattering values.}. This constrains $\delta V$.

\begin{figure}[t]
\centerline{\includegraphics[width=0.9\textwidth]{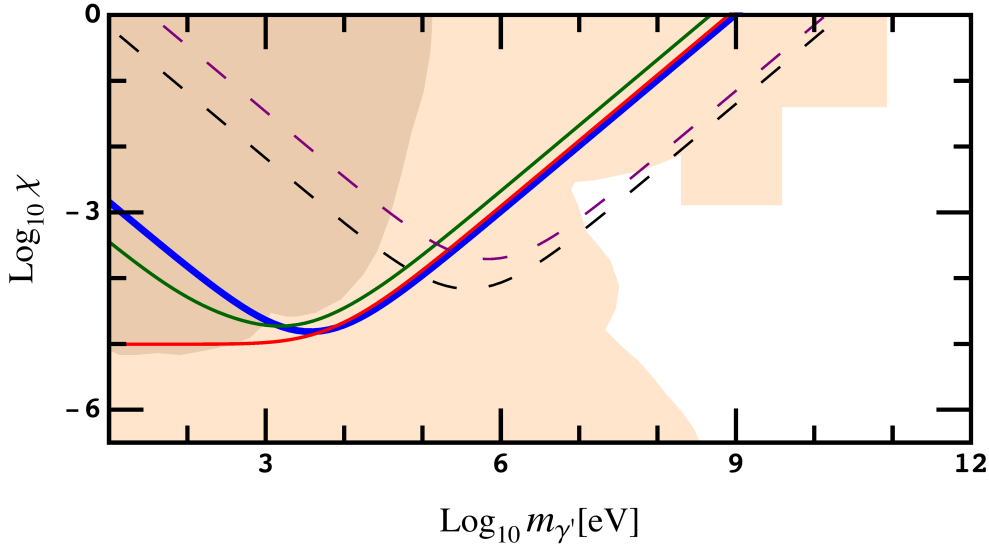}}
\caption{Bounds on hidden photons. The thick blue curve is from the $2s_{1/2} - 2p_{1/2}$ transition in atomic hydrogen. The red curve shows the naive bound from $1s_{1/2} - 2s_{1/2}$ in atomic hydrogen. It behaves incorrectly at small $m_{\gamma^{\prime}}$. The thin joined (green) curve is correctly renormalised by combining $1s_{1/2} - 2s_{1/2}$ with $2s_{1/2} - 8s_{1/2}$.
The colour filled regions are existing bounds. Those from pure Coulomb's law tests are the darker section (brown) \cite{Popovontheexp,Karshenboim:2010cg, Karshenboim:2010ck}.
The lower dashed curve (black) shows a speculative bound from $2s_{1/2}-2p_{1/2}$ in true muonium, using only theoretical values. The upper dashed curve (purple) uses experimental data from muonic hydrogen to form another speculative bound. Both speculative curves penetrate untested parameter space.
\label{condensed}}
\end{figure}

We find that a naive application of this method fails. Transitions between states with different values of the principal quantum number $n$ do not exhibit the correct drop off for small masses. For example the naive $1s_{1/2} - 2s_{1/2}$ bound from atomic hydrogen is plotted as the dotted red curve in Fig.~\ref{condensed}. 
We can understand the physical reasoning for this from Eq.~\eqref{Vintro}. At small masses our perturbation reduces to a term that has the form of a Coulomb potential, but with an extra factor $(1+\chi^2)$, which we have not absorbed into $\alpha$. In other words we have forgotten to properly (re-)normalise the coupling $\alpha$. We do this by treating $\alpha$ as a unknown instead of a constant, and using a second transition to solve for it. We can then produce properly renormalised bounds which are functions of two transitions and not one. Fig.~\ref{condensed} shows a correctly renormalised bound (thin, joined, green) using $1s_{1/2} - 2s_{1/2}$ and $2s_{1/2} - 8s_{1/2}$ transitions in atomic hydrogen.

Note that Lamb shift\footnote{A Lamb shift is defined by the energy difference between two states with the same $n$, $j$.} bounds renormalise trivially and can therefore be formed using only one measurement (see \cite{Jaeckel:2010xx}). The $2s_{1/2} - 2p_{1/2}$ bound from atomic hydrogen is the thick blue joined curve in \ref{condensed}. The bound drops off correctly for small masses.

Other transitions involving higher excited states in atomic hydrogen are considered. However these do not form good bounds, mainly due to high experimental uncertainties (see \cite{Jaeckel:2010xx}).

We then apply the method to other atomic systems. The idea is that other atomic systems may have advantages over atomic hydrogen. For example in pure QED systems like muonium and positronium, we can assume a point-like nucleus \cite{Jungmann:2004sa}. This eliminates large uncertainties from finite nuclear size effects. Smaller uncertainties strengthen our bounds, as can be seen from Eq. (\ref{1st order}). 
\par However in many cases theoretical uncertainties are larger. For example hadronic orbiting particles also interact with the nucleus via the strong interaction, which causes huge theoretical and experimental uncertainties (see \cite{Jaeckel:2010xx}). 
\par Also many of these atoms have larger reduced masses or smaller Bohr radii than atomic hydrogen, shifting our bounds to higher masses and towards the unexplored region.

All relevant transitions in hydrogen-like atoms were examined to see whether the advantages outweigh the disadvantages. We found overall that we could not improve upon our original atomic hydrogen bounds using existing data\footnote{This is actually a slight  simplification. The Lamb shift bound for the $Z = 2$ hydrogen-like ion is marginally stronger than the corresponding one for atomic hydrogen. This is because the $r_{p}$ anomaly causes a high level of theoretical uncertainty in atomic hydrogen and weakens the bound considerably. No such anomaly exists for measurements of alpha particle charge radius, and the helium-like hydrogen Lamb shift gives us a slightly stronger bound. However the general trend is for bounds to weaken as the nuclear charge $Z$ increases. We expect this trend to re-established as soon as the $r_{p}$ anomaly is resolved; the atomic hydrogen bound should then be the strongest.}.

Our best bounds (the light green and thick dark blue lines in Fig.~\ref{condensed}) do not penetrate new parameter space for hidden photons. However they do improve upon previous Coulomb's law tests (brown region  in \ref{condensed}) in the sense that they extend the excluded region to higher masses. This is a non-trivial improvement as Coulomb based bounds are especially clean and model independent \cite{Jaeckel:2010xx, Karshenboim:2010cg, Karshenboim:2010ck}. For example fixed target bounds assume a 100 $\%$ branching ratio for hidden photons to decay into charged standard model particles. If this assumption is wrong, then the bounds are weakened or possibly invalidated (see \cite{Jaeckel:2010xx}). 

Finally we investigate the discovery potential of future experiments. 

For example, a recent article suggests that true muonium ($\mu^{+}$ $\mu^{-}$) could be produced and studied in the near future. The reduced mass is $\sim$ 100 times greater than atomic hydrogen, and for a pure QED system we expect small theoretical errors.
Since no experimental data is available, we produce a speculative bound using an estimate of the theoretical error (lower dashed, black in Fig.~\ref{condensed}). This penetrates new parameter space, but one still needs to obtain a coherent experimental result.

The reduced mass of muonic atoms are $\sim$ 200 times larger than atomic hydrogen. In references \cite{Eides:2000xc,PhysRevA.53.2092,PhysRevA.60.3593,PhysRevA.71.032508,PhysRevA.71.022506,hypfinesplit,citeulike:7426442} the $2s_{1/2}^{F=1} - 2p_{3/2}^{F=2}$ difference in muonic hydrogen is calculated as a function or the proton radius $r_{p}$ . If we substitute in the most precise current value of $r_{p} = 0.8768(69)$ fm, from atomic spectra \cite{mohr-2007}, we obtain $E_{th} = -205.984(062)$ meV. The theoretical uncertainty alone is quite high. 
Moreover this also deviates from the recently measured experimental value of $-206.295000(3)$ meV \cite{citeulike:7426442} by around 5$\sigma$. This discrepancy is bad for producing bounds, but could be taken as a potential signal for new physics. We considered if the hidden photon could be used to explain this anomaly \cite{falkowskidark}. However this is ruled out by Lamb shift measurements in atomic hydrogen (see \cite{Jaeckel:2010xx}).

However we can form a speculative bound (upper dashed, purple in Fig.~\ref{condensed}) from just experimental uncertainty. This bound penetrates new parameter space.
If an independent and sufficiently precise value of $r_{p}$ could be determined -- consistent with the muonic hydrogen extraction --
this speculative bound could be turned into a real one.

\section{Conclusion}
\label{Conclusion}

We have used atomic spectroscopy of ordinary and exotic atoms to constrain deviations from Coulomb's law. A fully renormalised method was developed, which provides correctly shaped constraints for high and low masses.
We interpreted these constraints as bounds on hidden photons and found that pure Coulomb bounds were extended to higher masses. This is a non-trivial improvement as Coulomb based bounds are especially clean and model-independent, and provide complementary information to existing ones, which are often more model dependent.
We also find that new parameter space for hidden photons could be penetrated using future data from exotic atoms.

\section*{Acknowledgements}
The authors would like to thank A.~Lindner, J.~Redondo and A.~Ringwald for interesting discussions and useful comments.

\bibliographystyle{h-physrev5}
\bibliography{SBNP.bib}

\end{document}